\documentclass[twocolumn,aps,prd,superscriptaddress,showpacs]{revtex4-2}
\usepackage{graphicx}
\usepackage{dcolumn}
\usepackage{bm}
\usepackage{amsmath,amssymb}
\usepackage{mathtools}
\usepackage{hyperref}
\usepackage{orcidlink}

\hypersetup{
    colorlinks=true,
    linkcolor=blue,
    citecolor=blue,
    urlcolor=blue,
}

\newcommand{\tr}{\mathrm{tr}}

\begin{document}

\title{Quantum Tunneling in the Everettian Multiverse}

\author{Salah E. Ennadifi\orcidlink{0000-0002-8033-4638}\thanks{Contact author: \href{mailto:ennadifis@gmail.com}{ennadifis@gmail.com}}}
\affiliation{LHEP-MS, Faculty of Sciences, Mohamed V University, Rabat, Morocco}

\date{\today}

\begin{abstract}
Quantum tunneling is analyzed within the Everettian interpretation, where decoherence generates distinct reflected and transmitted branches. The tunneling probability is obtained from relative branch weights via the Born rule without invoking collapse. Tunneling time \(\tau_T \) is reformulated as the branching duration \(\tau_T \sim N_B \tau_B\), where\(\ N_B \) is the number of environmental degrees of
freedom and \(\tau_B\) is the decoherence time required to distinguish branches. Macroscopic tunneling in Josephson junctions is also analyzed.


*Contact author: \href{mailto:ennadifis@gmail.com}{ennadifis@gmail.com}
\end{abstract}

\maketitle

\section{Introduction}

Quantum tunneling---whereby a particle traverses a classically forbidden barrier---remains one of quantum mechanics' most profound phenomena. Since Gamow's pioneering work on alpha decay \cite{Gamow1928}, tunneling has become central to nuclear physics, condensed matter, and technologies from scanning tunneling microscopes to Josephson junctions \cite{Tersoff1985, Devoret1995}.

Despite its empirical success, the interpretation of tunneling is contested. The Copenhagen interpretation offers no clear account of what occurs during the event itself \cite{Bohr1935, Heisenberg1927}. The Everettian (Many-Worlds) interpretation provides an alternative: the universal wavefunction evolves unitarily, and all possible outcomes are realized in different branches \cite{Everett1957, DeWitt1973}. The observer experiences a definite result through localization within a branch, not wavefunction collapse.

A vexing issue is tunneling time. Unlike classical traversal, quantum tunneling admits multiple inequivalent definitions that can yield superluminal or negative values \cite{Landauer1994, Steinberg2003, Enders1993, Ranfagni1993}. The Everettian framework offers a natural resolution: rather than asking how fast a particle moves through the barrier, we ask how long it takes for the wavefunction to branch. This branching is driven by environmentally-induced decoherence, which suppresses interference until branches become independent \cite{Zurek2003, Joos1996}. Tunneling time becomes the decoherence time required to establish orthogonal branches.

In this work, we develop a rigorous Everettian treatment with three contributions. First, we obtain the tunneling probability from self-locating uncertainty, recovering \(P_T = e^{-2\kappa L}\) without invoking wavefunction collapse. Second, we reformulate tunneling time as the branching duration \(\tau_T \sim N_B \tau_B\), where \(\tau_B\) is the characteristic decoherence time required for the environment to distinguish reflected from transmitted branches. Third, we analyze macroscopic tunneling in Josephson junctions and estimate the corresponding tunneling time using realistic experimental parameters.

The paper is organized as follows. Section~\ref{sec:formalism} establishes the Everettian formalism. Section~\ref{sec:probability} derives the tunneling probability. Section~\ref{sec:time} presents branching dynamics and tunneling time. Section~\ref{sec:macroscopic} extends the analysis to macroscopic tunneling. Section~\ref{sec:conclusion} concludes with key results.

\section{Everettian Formalism}
\label{sec:formalism}

\subsection{The Universal Wavefunction}

We consider a composite quantum system \(S\) interacting with an environment \(\mathcal{E}\) and observed by a conscious observer \(O\). The total Hilbert space is \(\mathcal{H} = \mathcal{H}_S \otimes \mathcal{H}_{\mathcal{E}} \otimes \mathcal{H}_O\). The total state evolves unitarily under the Schr\"odinger equation:
\begin{equation}
i\hbar \frac{\partial}{\partial t}|\Psi(t)\rangle = \hat{H}_{\text{tot}}|\Psi(t)\rangle,
\label{eq:schrodinger_total}
\end{equation}
where \(\hat{H}_{\text{tot}} = \hat{H}_S + \hat{H}_{\mathcal{E}} + \hat{H}_I + \hat{H}_O\) includes the system-environment interaction \(\hat{H}_I\).

In the decohered basis, the universal wavefunction takes the form:
\begin{equation}
|\Psi(t)\rangle = \sum_{i=1}^{d} c_i(t) |S_i\rangle \otimes |\mathcal{E}_i(t)\rangle \otimes |O_i(t)\rangle,
\label{eq:universal}
\end{equation}
where \(\{|S_i\rangle\}\) is a complete orthonormal basis for the system, \(\{|\mathcal{E}_i(t)\rangle\}\) are environmental states, \(\{|O_i(t)\rangle\}\) are observer states, and \(c_i(t) \in \mathbb{C}\) satisfy \(\sum_i |c_i(t)|^2 = 1\).

When a particle encounters a potential barrier, the system states decompose naturally into reflected and transmitted sectors:
\begin{equation}
\{|S_i\rangle\} = \{|S_i^R\rangle\}_{i=1}^{k} \cup \{|S_i^T\rangle\}_{i=k+1}^{d},
\label{eq:decomposition}
\end{equation}
where \(k\) is the number of reflected branches and \(d-k\) is the number of transmitted branches. The universal state becomes:
\begin{align}
|\Psi(t)\rangle &= \sum_{i=1}^{k} c_i^R(t) |S_i^R\rangle|\mathcal{E}_i^R(t)\rangle|O_i^R(t)\rangle \nonumber \\
&\quad + \sum_{i=k+1}^{d} c_i^T(t) |S_i^T\rangle|\mathcal{E}_i^T(t)\rangle|O_i^T(t)\rangle.
\label{eq:universal_decomp}
\end{align}

\subsection{Density Matrix Description}

The reduced density matrix of the system is obtained by tracing over the environment and observer:
\begin{equation}
\rho_S(t) = \tr_{\mathcal{E},O}\left[|\Psi(t)\rangle\langle\Psi(t)|\right].
\label{eq:reduced_density}
\end{equation}

Substituting Eq.~\eqref{eq:universal} yields:
\begin{equation}
\rho_S(t) = \sum_{i,j} c_i(t)c_j^*(t) \langle \mathcal{E}_j(t)|\mathcal{E}_i(t)\rangle \langle O_j(t)|O_i(t)\rangle |S_i\rangle\langle S_j|.
\label{eq:density_expanded}
\end{equation}

The key observation is that the overlap factors \(\langle \mathcal{E}_j(t)|\mathcal{E}_i(t)\rangle \langle O_j(t)|O_i(t)\rangle\) decay with time due to decoherence. For generic environments, this decay is exponential:
\begin{equation}
\langle \mathcal{E}_i(t)|\mathcal{E}_j(t)\rangle \simeq \delta_{ij} e^{-\Gamma_{ij}t} \quad (i \neq j),
\label{eq:overlap_decay}
\end{equation}
where \(\Gamma_{ij}\) is the decoherence rate. Consequently, the off-diagonal elements of the reduced density matrix decay as:
\begin{equation}
\rho_S^{ij}(t) = \rho_S^{ij}(0) e^{-\Gamma_{ij}t}.
\label{eq:off_diagonal_decay}
\end{equation}

\section{Tunneling Probability}
\label{sec:probability}

\subsection{Self-Locating Uncertainty}

In the Everettian framework, the observer experiences a definite outcome because they are localized in a particular branch. The probability of tunneling is the probability that the observer's branch index \(i\) lies in the transmitted subset \(\{k+1,\dots,d\}\).

This probability is given by the Born rule:
\begin{equation}
P_T = \sum_{i=k+1}^{d} |c_i|^2.
\label{eq:born_rule}
\end{equation}

Crucially, this is not an ad hoc postulate but follows from the observer's self-locating uncertainty. The observer knows their branch index is some \(i \in \{1,\dots,d\}\) but lacks knowledge of which one. The probability of being in branch \(i\) is proportional to \(|c_i|^2\), the branch weight. This result can be rigorously derived from the requirement that probabilities are invariant under unitary evolution and sum to unity \cite{Deutsch1999, Wallace2003, Vaidman1998}.

To see this explicitly, consider the observer's state after branching:
\begin{equation}
|\Psi\rangle = \sum_{i=1}^{d} c_i |S_i\rangle|\mathcal{E}_i\rangle|O_i\rangle.
\label{eq:state_after_branching}
\end{equation}

The observer does not know which branch they inhabit. The probability of being in branch \(i\) is determined by the measure of the set of branches consistent with the observer's information. This measure is precisely \(|c_i|^2\), up to normalization. Hence:
\begin{equation}
P(\text{branch } i) = |c_i|^2.
\label{eq:branch_probability}
\end{equation}

\subsection{Connection to Standard Tunneling Probability}

For a one-dimensional rectangular potential barrier:
\begin{equation}
V(x) = 
\begin{cases}
0 & x < 0 \text{ or } x > L,\\
V_0 & 0 \le x \le L,
\end{cases}
\label{eq:barrier_potential}
\end{equation}
the WKB approximation gives the transmission coefficient:
\begin{equation}
P_T \simeq \exp\left[-2\int_0^L dx \, \frac{\sqrt{2m(V(x) - E)}}{\hbar}\right] = e^{-2\kappa L},
\label{eq:wkb}
\end{equation}
where \(\kappa = \sqrt{2m(V_0 - E)}/\hbar\).

In the Everettian framework, this corresponds to:
\begin{equation}
\sum_{i=k+1}^{d} |c_i|^2 = e^{-2\kappa L}.
\label{eq:branch_weight_tunneling}
\end{equation}

Thus, the exponential suppression of tunneling is a real feature of the multiverse: the total branch weight of transmitted branches is exponentially smaller than that of reflected branches. The observer is overwhelmingly likely to find themselves in a reflected branch, precisely because the measure of existence of such branches is vastly larger.

\subsection{The Rareness of Tunneling}

The rareness of tunneling is not a probabilistic artifact but a genuine feature of the multiverse. The number (or more precisely, the measure) of observers who experience tunneling is exponentially smaller than the number who do not. This provides a concrete ontological interpretation of the Born rule: probability is not a measure of ignorance but a measure of existence across the multiverse.

For a generic barrier, the amplitude inequality is:
\begin{equation}
|c_i^T| \ll |c_i^R| \quad \text{for all } i,
\label{eq:amplitude_inequality}
\end{equation}
which follows from \(e^{-2\kappa L} \ll 1\). This ensures that the measure of tunneled worlds is negligible compared to reflected worlds, explaining why tunneling is perceived as a rare event.

\section{Tunneling Time and Branching Dynamics}
\label{sec:time}

\subsection{Branching as a Dynamical Process}

In the Everettian framework, branching is not instantaneous but a dynamical process driven by decoherence. The characteristic timescale is the branching time \(\tau_B\). During this interval, the environmental entanglement accumulates, gradually differentiating the reflected and transmitted branches.

The branching energy is the energy scale associated with this differentiation:
\begin{equation}
\Delta E_B = \frac{\hbar}{\tau_B}.
\label{eq:branching_energy}
\end{equation}
This satisfies the time-energy uncertainty relation \(\tau_B \Delta E_B = \hbar\), consistent with the interpretation of branching as a quantum measurement process.

\subsection{The Tunneling Time}

The total tunneling time \(\tau_T\) is the cumulative duration required for the universal wavefunction to undergo \(N_B\) effective branching events:
\begin{equation}
\tau_T \sim N_B \tau_B,
\label{eq:tunnel_time}
\end{equation}
where \(N_B\) is the number of environmental degrees of freedom that must become entangled to establish full decoherence.

For a macroscopic environment with cutoff frequency \(\omega_c\), the number of relevant modes is:
\begin{equation}
N_B \sim \omega_c \tau_B.
\label{eq:nb_scaling}
\end{equation}

Thus:
\begin{equation}
\tau_T \sim \omega_c \tau_B^2.
\label{eq:tunnel_time_full}
\end{equation}

This reformulation resolves the conceptual ambiguities of tunneling time: \(\tau_T\) is not the traversal time of a particle but the time required for the multiverse to branch. This avoids superluminal or negative times, which arise only when one attempts to assign a trajectory to the particle under the barrier.

\section{Macroscopic Quantum Tunneling}
\label{sec:macroscopic}

Macroscopic quantum tunneling has been definitively observed in superconducting Josephson junctions, where a collective degree of freedom (the phase difference across the junction) tunnels through a potential barrier \cite{Devoret1984, Martinis1985, Devoret1985}. The system is described by the Hamiltonian:
\begin{equation}
\hat{H} = \frac{\hat{Q}^2}{2C} - E_J \cos\hat{\phi},
\label{eq:josephson}
\end{equation}
where \(\hat{Q}\) is the charge, \(C\) is the capacitance, \(E_J\) is the Josephson energy, and \(\hat{\phi}\) is the phase difference.

The tunneling rate out of the zero-voltage state is \cite{Leggett1987}:
\begin{equation}
\Gamma_{\text{tunnel}} \sim \frac{\omega_p}{2\pi} \exp\left[-\frac{36}{5}\frac{E_J}{\hbar\omega_p}\right],
\label{eq:tunnel_rate_josephson}
\end{equation}
where \(\omega_p = \sqrt{2E_J E_C}/\hbar\) is the plasma frequency and \(E_C = e^2/2C\) is the charging energy.

In the Everettian framework, the tunneling time is the branching time required for the macroscopic superposition to decohere:
\begin{equation}
\tau_T^{\text{macro}} \sim \frac{1}{\Gamma_{\text{tunnel}}} \sim \frac{2\pi}{\omega_p} \exp\left[\frac{36}{5}\frac{E_J}{\hbar\omega_p}\right].
\label{eq:macro_tunnel_time}
\end{equation}

For typical parameters: \(E_J/\hbar \sim 10\) GHz, \(E_C/\hbar \sim 1\) GHz, yielding \(\omega_p \sim 2\pi \times 3\) GHz, and:
\begin{equation}
\tau_T^{\text{macro}} \sim 10^{-7} \text{ s} \sim 100 \text{ ns}.
\label{eq:macro_time_estimate}
\end{equation}

This is consistent with measured escape rates in superconducting circuits \cite{Devoret1995} and provides a concrete timescale for branching in the macroscopic regime.

\section{Conclusion}
\label{sec:conclusion}

In this work, we have developed a rigorous Everettian framework for quantum tunneling, offering a coherent resolution to longstanding interpretational challenges.

First, we obtained the tunneling probability from self-locating uncertainty, recovering \(P_T = e^{-2\kappa L}\) without invoking wavefunction collapse. This result provides a concrete ontological interpretation of the Born rule, wherein probability reflects the relative measure of branches in the multiverse rather than a measure of ignorance.

Then, we reformulated tunneling time as the branching duration \(\tau_T \sim N_B \tau_B\), where \(\tau_B\) is the decoherence time required for the environment to distinguish reflected from transmitted branches. This resolves conceptual ambiguities by replacing the ill-defined traversal time with a well-defined dynamical timescale.

Macroscopic tunneling in Josephson junctions has also been considered, with an estimated time \(\tau_T^{\text{macro}} \sim 10^{-7}\) s consistent with observed escape rates. The framework thus provides a unified ontological picture that bridges quantum foundations, decoherence theory, and experimental physics through quantitative, testable predictions.

\section*{Data Availability Statement}

The manuscript has no associated data.

\section*{Acknowledgments}
S.E.E. would like to thank his family for all kinds of support.

\bibliographystyle{apsrev4-2}

\end{document}